# Cytoplasmic flows as signatures for the mechanics of mitotic positioning

Ehssan Nazockdast[(1)], Abtin Rahimian[(1)], Daniel Needleman[(2)], and Michael Shelley[(1)]

[(1)]Courant Institute of Mathematical Sciences, New York University
[(2)]School of Engineering and Applied Sciences, Harvard University

## Abstract

The proper positioning of the mitotic spindle is crucial for asymmetric cell division and generating cell diversity during development. Proper position in the single-cell embryo of *Caenorhabditis elegans* is achieved initially by the migration and rotation of the pronuclear complex (PNC) and its two associated centrosomal arrays of microtubules (MTs). We present here the first systematic theoretical study of how these $\mathcal{O}(1000)$ centrosomal microtubules (MTs) interact through the immersing cytoplasm, the cell periphery and PNC, and with each other, to achieve proper position. This study is made possible through our development of a highly efficient and parallelized computational framework that accounts explicitly for long-ranged hydrodynamic interactions (HIs) between the MTs, while also capturing their flexibility, dynamic instability, and interactions with molecular motors and boundaries. First, we show through direct simulation that previous estimates of the PNC drag coefficient, based on either ignoring or partially including HIs, lead to misprediction of the active forces and time-scales of migration by an order of magnitude. We then directly study the dynamics of PNC migration under various force-transduction models, including the pushing or pulling of MTs at the cortex, and the pulling of MTs by cytoplasmically-bound force generators. While achieving proper position and orientation on physiologically reasonable time-scales does not uniquely choose a model, we find that each model produces a different signature in its induced cytoplasmic flow and MT conformations. We suggest then that cytoplasmic flows and MT conformations can be used to differentiate between mechanisms and to determine their contribution to the migration process. Since these flow signatures are generic features of each model, they can used to investigate the presence of active processes in other stages of the cell division and other organisms.

## 1 Introduction

The cytoskeleton is an assembly of microscopic filaments and molecular motors, and the machinery for performing many vital cellular processes [Fletcher and Mullins, 2010; Howard, 2001]. Despite its critical role in the progression of life, important aspects of cytoskeletal mechanics remain poorly understood, particularly how the interactions of microscopic cytoskeletal elements — filaments and motor-proteins — with each other and the cellular cytoplasm relate to observed larger-scale behavior. Microrheological measurements do provide some useful estimates of the mechanical responses of complex materials, but depend upon specific microscopic models for interpretation. Such microscopic models are substantially missing for active and heterogeneous biological systems. Most theoretical models of the mechanics of cytoskeleton are also limited to continuum theories with limited linkages to the interactions of fibers and motors at the microscopic level [Bausch et al., 1999; Fabry et al., 2001; Moeendarbary et al., 2013].

Dynamic simulation can be a powerful tool for studying these relationships. For this purpose we have developed versatile and highly efficient numerical tools for studying the dynamics of active and flexible filaments in cellular assemblies. This is, to our knowledge, the first attempt to incorporate many-body interactions between MTs and other intracellular bodies with the cytoplasmic fluid, while also accounting for the mechanical flexibility of MTs, their dynamic instability, and interactions with motor proteins. These fluid-structure interactions are commonly referred to as *hydrodynamic interactions* (HIs). By accounting for HIs, we can compute the large-scale cytoplasmic flows generated by the movements of MTs and other immersed bodies within cell cortex.

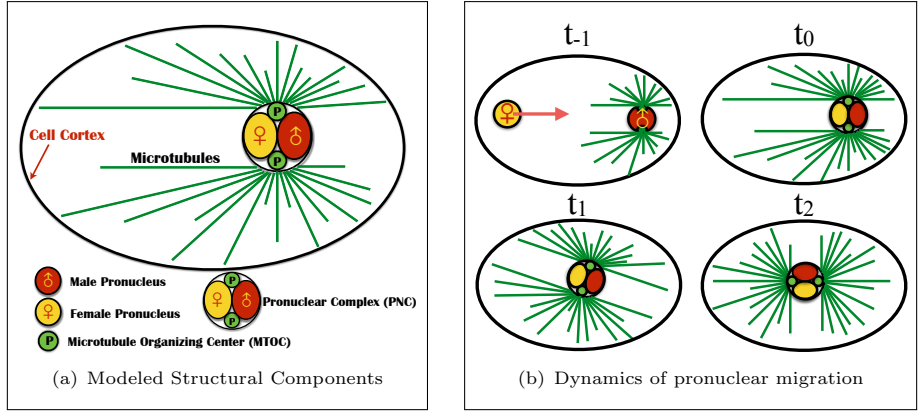

**Figure 1:** *A schematic illustrating the modeled structural components and dynamics of pronuclear migration in the single-cell C. elegans embryo. (a) the structural components. The pronuclear complex — here modeled as a rigid sphere — contains the male (red) and female (yellow) pronuclei, and is attached to two arrays of MTs (green lines) that polymerize from two centrosomes (green bodies). These structures are immersed in the cellular cytoplasm and confined within an ellipsoidal eggshell. (b) The dynamics of pronuclear migration and positioning. At $t = t_{-1}$, the female pronucleus moves to the posterior to combine with the male pronucleus ($t = t_0$) to form the pronuclear complex (PNC). This initial period of female nuclear migration is not modeled here [Payne et al., 2003]. Between $t = t_0$ and $t_1$ the PNC moves anterior-wise to the center while rotating into "proper position" with the centrosomal axis along the anterior-posterior (AP) axis ($t = t_2$).*

At the core of our computational approach are specialized methods for simulating the dynamics of microscopic fibers and other structures that interact in a Newtonian fluid. Unlike other approaches to computing cellular flows (see, e.g., [Niwayama et al., 2011; Shinar et al., 2011]) ours is based upon boundary integral methods that express solutions to the Stokes equations in terms of singular integrals over all immersed surfaces [Power and Miranda, 1987; Pozrikidis, 1992]. This reduces solving the 3D fluid-structure problem to solving 2D singular integral equations for surface densities. For MTs, which are very slender, this formulation is further simplified using slender-body asymptotics which reduces the 2D surface integrals to 1D integrals over MT center-lines. This latter step also gives a natural framework for incorporating the elastic mechanics of MTs and for implementing implicit time-stepping methods to reduce temporal stiffness. After immersed and bounding surfaces, and MT center-lines are discretized and quadrature schemes applied, the simultaneous solution of all integral equations is accomplished using a preconditioned GMRES iteration [Saad and Schultz, 1986] where requisite matrix-vector multiplications are done in only $\mathcal{O}(N)$ operations, with $N$ the number of surface unknowns, using generalized-kernel fast multipole methods [Ying et al., 2004]. Proper preconditioning insures that solution is found in only a small number of iterations, independently of the number of discretization points. The end result is a highly efficient numerical approach for evolving geometrically complex structures of flexible MTs and other structures, while accounting for all HIs amongst them, at a computational cost of $\mathcal{O}(N)$ operations per time-step, and without severe time-step restrictions [Nazockdast et al., 2015; Tornberg and Shelley, 2004]. Given that we simulate $\mathcal{O}(100-1000)$ MTs, the number of discretization points is dominated by those resolving MTs, not other structures, and hence the computational cost essentially scales linearly with the number of MTs.

Here we apply these methods to study pronuclear migration during the first cell division of *C. elegans* embryo; see the schematic in Fig. 1. Proper positioning of the mitotic spindle is indispensable to the successful segregation of chromosomes and to the generation of cell diversity in early development [Cowan and Hyman, 2004]. Prior to mitosis and after fertilization, the female pronucleus migrates towards ($t = t_{-1}$) and meets the male pronucleus ($t = t_0$), and its associated centrosomal array of MTs, at the cell posterior to form the pronuclear complex (PNC). The PNC then moves towards the cell center ($t = t_1$; centering) and rotates 90 degrees ($t = t_2$) to align the axis between its two associated centrosomes with the cell's anterior-posterior (AP) axis. Afterwards, the mitotic spindle forms ($t > t_2$).

Hydrodynamic interactions arise from several features of the PNC migration. Centering and rotation of the PNC will push and rotate the cytoplasm, as will the centrosomal MT arrays. Hence, each structure moves against the backdrop of flows produced by the other. Further, since the PNC and its associated MT arrays are on the scale of the cell itself, the confinement of the cell will have a very strong effect on the nature of the cytoplasmic flows. Finally, as we shall show here, the particular mechanisms of force transduction that position the PNC can have a first-order effect on these flows. To establish the importance of hydrodynamic interactions on PNC migration, we perform the numerical experiment of pulling the PNC, and its centrosomal MT array, with an external force from the posterior to the center of the cell and aligning it with AP axis with an external torque. We then compare the computed translational and rotational drag on the PNC/MT-array complex with other estimates in the literature including those based on the radius of the aster formed by the MTs [Kimura and Onami, 2005, 2007; Reinsch and Gönczy, 1998], those using a local drag model [Nedelec and Foethke, 2007], and with our own previous study that ignored the MT drag altogether (though not PNC drag or the effect of confinement) [Shinar et al., 2011]. We find in each case that these estimates drastically over- or under-estimate this drag. We also find that confinement has a much stronger effect on translational motion than on rotational.

Three types of mechanisms have been proposed for pronuclear migration. In *cortical pushing* models, the growth of astral MTs against the cell periphery induces repulsive forces on MTs that move the complex away from the periphery and thus opens space for further polymerization or which stop the growth reaction altogether [Reinsch and Gönczy, 1998]. In a *cortical pulling* model, MTs impinging upon the cortex are pulled upon by dynein motors that are attached to the plasma membrane, in particular by association to the protein complex formed by the G$\alpha$ subunits, GPR-1/2, and LIN-5. An asymmetric distribution of PAR and LET-99 proteins on the cortex in prophase then produces an asymmetric association of dyneins with the protein complex, and larger anterior pulling forces on the anterior, and so the pronuclear complex moves towards that direction [Goulding et al., 2007; Grill et al., 2001; Kimura and Onami, 2007; Labbé et al., 2004; McNally, 2013; Siller and Doe, 2009; Tsou et al., 2002]. This asymmetry is reversed in anaphase with larger forces on the posterior which leads to asymmetric cell division in *C. elegans* [Kimura and Onami, 2007]. In a *cytoplasmic pulling* model, forces are supplied by cargo-carrying dyneins attached on MTs and walking towards the centrosomes [Kimura and Onami, 2007]. As a consequence of Newton's third law the force applied by dynein on MTs is equal and opposite to the force required to move the cargo through the cytoplasm [Shinar et al., 2011]. Since longer MTs carry more dyneins and produce larger pulling forces, the PNC moves in the direction of longer MTs, that is, anterior-wise.

Within our computational framework, we study the mechanics and dynamics of pronuclear migration using simple biophysical models of these various mechanisms while accounting for the detailed interactions of the MTs and other bodies through the cytoplasm. We show that the cortical and cytoplasmic pulling mechanisms, as well as one variation of cortical pushing model, can properly center and rotate the PNC. We also demonstrate that each mechanism produces its own *fingerprint* in the generated cytoplasmic flows and MT conformations which can be used to differentiate between them. These flow signatures are generic features of each mechanism and should not depend on the details of its biochemical regulation and molecular pathways. Specifically, we show that the cytoplasmic flow generated in the cortical pulling model is that arises from pushing a porous object with an external force. Little deformation of the MTs is observed as they are typically under an extensional tension from the force generators on the cortex. In the cortical pushing models, the cytoplasmic flow is the superposition of the driven porous object flow, and that produced by large MT deformations induced by compressive or bending loads at the periphery. These flows are different in our two variations of this model and are directly related to the conformations of the MTs (and to the lack of PNC rotation in one case). Finally we demonstrate that the flow induced by a cytoplasmic pulling model is fundamentally different as it can be interpreted as a porous structure that is moved by internal force generators, with its early time flows in the class of self-propelled Puller particles [Saintillan and Shelley,

2013].

While this study is limited to the pronuclear migration process in *C. elegans* embryo, the active mechanisms discussed here, including the polymerization forces or forces from cortically- or cytoplasmically-bound dyneins, are utilized in other stages of cell division and in other organisms [Howard, 2001]. Thus, the generic features of these mechanical models including their flow signatures can be useful in identifying or differentiating between these force transduction mechanisms in other instances.

## 2 The importance of hydrodynamic interactions

The mechanical role of astral MTs in all three centering mechanisms is to transfer the force applied either on their plus-ends (by cortical pushing or pulling), or along their lengths (by cytoplasmic pulling), to their minus-ends anchored in the centrosomes, themselves attached to the PNC. The induced velocity of the PNC is related to the net force on it by the instantaneous hydrodynamic resistance tensor of the entire structure composed of the PNC and its attached astral MTs, $\mathbf{f} = \mathbf{R}_{pnc} \cdot \mathbf{v}_{pnc}$ where, $\mathbf{R}_{pnc}$ is the $3 \times 3$ translational resistance tensor. In motion the MTs themselves can also have considerable drag, relative to the PNC, since in Stokes flow the resistance of an object scales roughly with its longest dimension. To see this, consider an MT of length $10\mu m$, which is about the diameter of the PNC and on the scale of the average MT length, being moved transversely to itself at a constant speed. Slender body theory can then be used to estimate the drag force on the MT, while the classical Stokes' formula estimates the drag force on a spherical PNC moving at the same speed. The ratio of these two drags (MT to PNC) is given by the formula $(4/3)\ln(\epsilon^{-2}e) \approx 0.25$ where $\epsilon = 1/400$ is the aspect ratio of the MT. Hence, despite having a diameter of only $20nm$ the MT nonetheless has 25% of the PNC's drag. Thus, MT drag during positioning the PNC can be substantial. Note that the flow generated by one MT strongly affects the motion of others so long as the average distance between them is within their average length. This is indeed the case for the astral MTs which argues for including HIs between them.

To first study the effect of HIs on PNC positioning, we perform model numerical experiments by first pulling the PNC (of diameter $10\mu m$) and its associated centrosomal arrays with a given external force from the cell posterior to its center and then rotating the PNC to proper position by applying an external torque. Here the cell is ellipsoidal in shape with major AP-axis length of $50\mu m$ and minor axis length of $30\mu m$. In these simulations MTs undergo dynamic instability (grow, shrink, undergo catastrophe and rescue [Desai and Mitchison, 1997] which is modeled via Poisson processes as in [Kimura and Onami, 2005]. The dynamic instability process is the only source of stochastic fluctuations in this model. Further, for the parameters used here an MT reaches an average length of $8\mu m$ before depolymerizing and so many MTs will reach the cell periphery during their growth. The observations of Srayko et al. [2005] suggest that during migration there are at least 600 centrosomal MTs. In investigating the effect of MT number on PNC dynamics, we will take $N_{MT} = 1200$ as the maximum number of MTs in our simulations. The dimensionless translational drag along the AP axis, $\hat{\gamma} = F/(6\pi\eta U_{pnc}a_{pnc})$, and the dimensionless rotational drag, $\hat{\gamma}_R = T/(8\pi\eta\omega_{pnc}a_{pnc}^3)$, are computed using values of translational speed, $U_{pnc}$, and angular velocity, $\omega_{pnc}$, measured from the simulations.

A number of studies [Kimura and Onami, 2005, 2007; Reinsch and Gönczy, 1998] have used the Stokes drag law, $F = 6\pi\mu U r$, to approximate the PNC/MT-array drag force where $r$ is an average radius of the aster formed by the MTs. Brownian Dynamics simulation techniques for modeling cellular assemblies, most notably *Cytosim* [Nedelec and Foethke, 2007], neglect HIs between the MTs, as well as between and with other immersed structures, but do include a local drag on all objects, and in particular on the MTs. In an approximation made in our own previous work, the drag upon MTs and the HIs between them were neglected while the fluid forces generated by motor-protein transport along MTs was accounted for, as were the flows generated by these forces, the PNC, and the cortex. Here we show that a complete accounting of HIs between MTs, the PNC, and the cell periphery gives drag coefficients an order of magnitude larger than the values used in [Kimura and Onami, 2005, 2007; Shinar et al., 2011] and an order of magnitude smaller than the predictions of the local drag model [Nedelec and Foethke, 2007].

To make these comparisons, we consider three different conditions or models. In model (1), we neglect all HIs, and calculate drag on MTs using a local slender body drag formula [Tornberg and Shelley, 2004], and drag on the PNC using Stokes' drag formula. In model (2), we include HIs amongst the MTs and with the PNC, but neglect the backflow generated by the cell periphery. These are both open flow approximations since HIs with the cell periphery are neglected. Still, its confining presence is partially maintained by having MTs depolymerize upon reaching it. In model (3), we include all HIs.

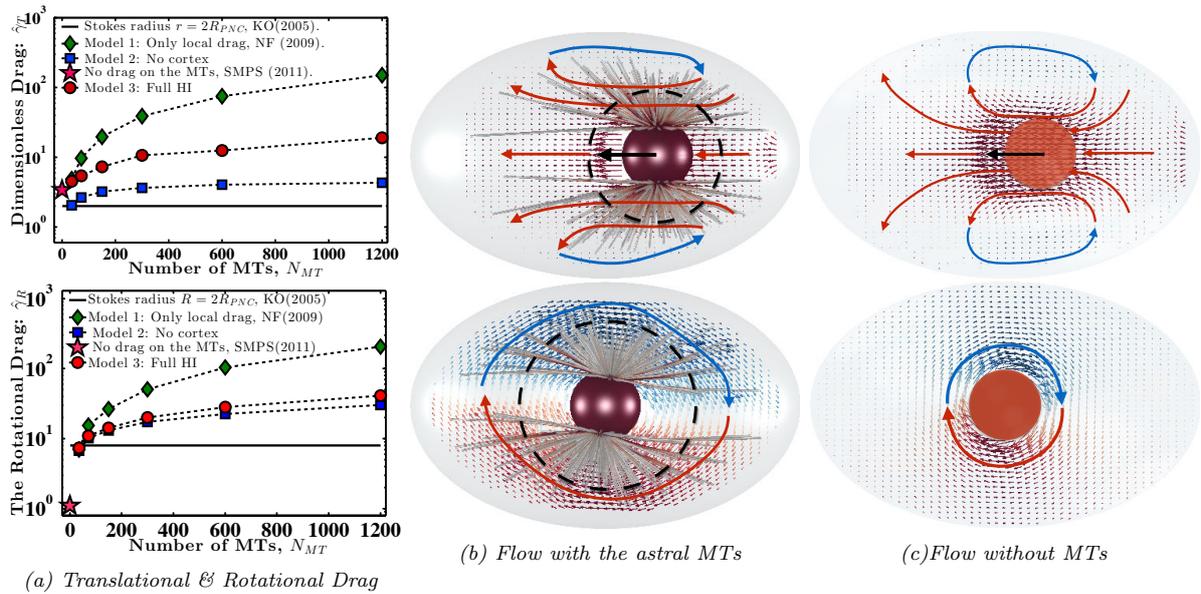

**Figure 2:** *(a) The dimensionless translational, $\hat{\gamma}_T = \gamma_T/\gamma_T^0$, and rotational, $\hat{\gamma}_R = \gamma_R/\gamma_R^0$ drag coefficients vs. the number of MTs ($N_{MT}$) where $\gamma_T^0 = 6\pi\eta R_{pnc}$ and $\gamma_R^0 = 8\pi\eta R_{pnc}^3$, are the translational and rotation drag coefficients, respectively, of a rigid spherical PNC in the absence of MTs. For a PNC with centrosomal MT arrays, (b) upper panel shows the cytoplasmic flow for PNC translation, and the lower panel shows that for rotation. The upper and lower panels of (c) are the same conditions, respectively, but without centrosomal MTs.*

Figure 2 shows the variation of measured translational and rotational drag coefficients with the number of MTs for all models (1, 2, and 3). The black reference lines at $\hat{\gamma}_T = 2$ and $\hat{\gamma}_R = 8$ are estimates that assume that PNC drag can be modeled by from Stokes formula using an effective sphere of radius $R = 2R_{pnc}$. This approximation was used by [Kimura and Onami, 2005, 2007] in their modeling of pronuclear rotation and centering.

As expected, for model (1), having no HIs between MTs, both drag coefficients increase linearly with $N_{MT}$, showing no saturation. For model (2), $\hat{\gamma}_T$ and $\hat{\gamma}_R$ increase monotonically with $N_{MT}$, with $\hat{\gamma}_T$ saturating as $N_{MT} \to \infty$. In another study, we show that for the range of forces used in this numerical experiment the Brinkman model of porous media flow gives an accurate description of the fluid flow for model (2) [Nazockdast et al., 2015]. In that continuum approximation, this problem corresponds to measuring the drag coefficient of a sphere (the PNC) surrounded by a co-moving porous layer that abuts the cell periphery. The HIs between the MTs reduce the convective penetration of fluid into the porous layer and creates an effectively larger object moving through the fluid. In the limit $N_{MT} \to \infty$ the flow cannot penetrate the porous layer, which fills the cell volume, and hence both drag coefficients saturates. The lack of a reaction flow at the cell periphery through the no-slip condition — missing in model (2) — means that there are cytoplasmic flows passing through the cell periphery.

When confinement-induced reaction flows are included by accounting for HIs with the periphery, model (3), the drag coefficients again show monotonic increase with $N_{MT}$. However, for $N_{MT} = 1200$, $\hat{\gamma}_T$ is 6-fold larger than that of model (2), is several-fold smaller than model (1), and is 6-fold smaller than model (3) with $N_{MT} = 0$ (the red star), which corresponds roughly to our previous modeling [Shinar et al., 2011] of pronuclear migration in which MT drag was not included.

Figure 2a (lower) shows that the rotational drag coefficient qualitatively follows the same trend as the translational coefficient with one major difference: the confinement has a much smaller effect on the dynamics of rotational as compared with translational motion.

This simple example demonstrate that the HI cannot be coarse-grained through a single modified viscosity since it does not have the same dynamical effect on both rotational and translational motions. Instead our simulations show generally that the entire cytoplasm-filled MT array acts as a porous medium whose permeability decreases with an increasing number of MTs, thus giving an increased effective size of the PNC in response to an applied force (at least in the absence of force generators within the centrosomal array producing active flows). This effect is nicely visualized by comparing the generated cytoplasmic flows in the presence and absence of MTs, shown in Figs. 2b and 2c, respectively (using $N_{MT} = 600$). Comparing the top panels shows that for translational flow, the presence of the centrosomal array does not qualitatively change the generated cytoplasmic flow, but does reduce the size of the reversing flow zone. This is effectively an increase in the hydrodynamic radius of the PNC, illustrated by the dashed circle. This increase in hydrodynamic radius is also apparent in rotational motion (bottom panels) by noting that the magnitude of the bulk velocity decays much more slowly away from the PNC when the astral MTs are present. In the limit of $N_{MT} \to \infty$, the effective hydrodynamic dimensions of PNC approaches the size of the cell periphery and we expect $\hat{\gamma} \to \infty$ due to the no-slip condition.

To sum up, this study shows that the drag on the astral MTs and confinement of the cytoplasm play a large role in determining the resistance to motion of the entire structure, and that their effects cannot be easily parametrized into simple reduced models.

## 3 Three mechanisms of pronuclear positioning and their cytoplasmic flows

Here we use the induced cytoplasmic flows, and MT conformation as generic features specific to each mechanism to differentiate between them. The motivation is that different mechanisms of force transduction will exert different forces upon MTs and hence MTs should develop different shapes in response. Similarly, PNC and MT-array motion from these different mechanisms should also be associated with generically different cytoplasmic flows.

To start, within our framework we have instantiated a *cortical pulling model* due to [Kimura and Onami, 2007] used in their study of pronuclear migration. Figure 3a shows a schematic of the model which was motivated by experimental observations of [Park and Rose, 2008]. In this model, pulling forces on centrosomal MTs is based upon their asymmetric attachment to cortically bound dyneins whose activation probability is inhomogeneously distributed along the cortex with the greatest probability of attachment at the posterior pole, and the least to the immediate posterior of the mid-plane. Attached MTs are pulled upon and simultaneously depolymerized. See §A for details. Snapshots from the simulation shown in Fig. 3d demonstrate that this mechanism leads to centering and rotation to proper position. In this simulation, PNC translation and rotation are in temporal register and proper position is achieved on a reasonable time-scale; see Fig. 3c. Since the cortically-bound force generators in this model put the MTs under an extensional load, MT deformations are small and they remain relatively straight; see Fig. 3b.

Since the MTs are hardly deformed in this model, the entire PNC/MT-array complex moves essentially as a single rigid body and generates a cytoplasmic flows similar to that of a porous body moving within a confined geometry. As Fig. 3d (top) illustrates, at early times the cytoplasmic velocity field is dominated by translation and bears a striking resemblance to that shown in Fig. 2b (top) where the PNC/MT-array is translated by an external force. As the PNC centers and translation slows, the cytoplasmic flow becomes primarily rotational, see Fig. 3d (middle), and the rotational velocity of the PNC is stronger than its translational velocity and the cytoplasmic flow is similar to that seen in Fig. 2b (middle), where the centered PNC/MT-array is rotated by an external torque. At late times, e.g. Fig. 3d (bottom), the cytoplasmic flows are weak and arise from fluctuations in PNC position due to the stochastic attachment and detachment of MTs from the cortex. Due to the MTs remaining straight under their extensive loading from cortical force generators, the cytoplasmic velocity fields arise almost entirely from the translations and rotations of the PNC/MT-array complex.

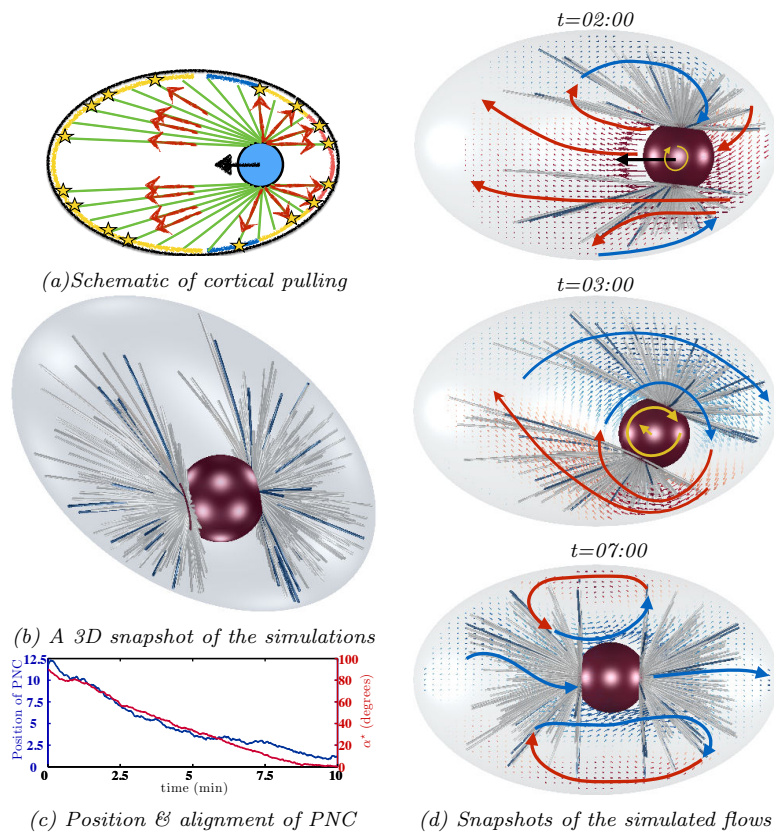

**Figure 3:** *(a) A schematic presentation of the cortical pulling model. The blue, yellow and red strips on the cortex correspond to the lowest, average and highest density, respectively, of active cortical dyneins. See §A for details. (b) A 3D snapshot of the simulation results. The fibers are color-coded with respect to the local tension; red, blue and white colors denote compressional, extensional, and no forces respectively. The same coloring scheme is applied to the rest of the figures. (c) The variations of the PNC position and angle between the intercentrosomal axis and the AP-axis with time. (d) Snapshots of the generated flows superimposed on the shape of MTs at different stages of pronuclear migration. Numbers denote time in minutes:seconds after the nuclei meeting. The results are projected onto the xy-plane to aid visualization, where the AP-axis is the x-axis and the intercentrosomal axis is initially aligned with the y-axis. Note that both the shape of the MTs and the flow are defined in three dimensions.*

For the parameters used in this simulation (motor attachment distribution and number density, MT attachment and detachment rates, etc.) we find that the PNC robustly finds proper position. These chosen parameters were physiologically reasonable but also narrowly constrained. Other seemingly reasonable choices of these parameters can lead to lack of centering. More details are given in §A.

Next we consider two variations of *cortical pushing models*. In these models the cortex exerts pushing forces on the plus-ends of MTs which reach and polymerize at the cell cortex. The magnitude of this force is such that it provides space — either by deforming the MT or pushing it away — for polymerization of MTs to continue. In the *free-sliding* submodel, illustrated schematically in Fig. 4a, these cortical repulsive forces are applied only in the normal direction to the cortex (pointing inwards) so that MT plus-ends cannot penetrate the boundary. MTs are nonetheless allowed to grow or slide freely tangentially. In the *no-sliding* submodel, illustrated schematically in Fig. 5a, we constrain both the position of MT plus-ends that reach cortex to remain fixed in position (trapped) on the cortex as long as they are in the growing state; See §A for details. In both submodels there is a force *vs.* polymerization velocity relationship imposed [Dogterom and Yurke, 1997; van Doorn et al., 2000], and introducing a sliding friction would give a continuum of cortical pushing models bridging these two extremes. These two models can induce different cytoplasmic flows near the cortex. In the free-sliding model the growth through polymerization at MT plus-ends will induce no cytoplasmic flow if the MT is aligned tangentially to the cortex at its plus-end. On the other hand, when the MT plus-end is pinned at the cortex then polymerization forces push newly formed microtubular material away from the cortex. This outward flux creates a flow by dragging cytoplasm from the cell periphery into the cell volume.

Figures 4d and 5d show snapshots from long-time simulations of both submodels of the cortical pushing model. In both cases the cytoplasmic motions at very early times are dominated by translation of the PNC/MT-array and are grossly similar to those seen in Fig. 2b (top). At early times, as before, the cytoplasmic velocity is dominated by translation and is similar to that seen in Fig. 2b (top). The two models' subsequent dynamics are much different, with the most consequential difference being that, while both models yield centration of the PNC/MT-array complex, only the no-sliding model shows rotation of the intercentrosomal axis to proper position (*cf.* the bottom panels of Fig. 4d with Fig. 5d, and Fig. 4c with Fig. 5c).

Unlike the cortical pulling model, both cortical pushing submodels show substantial, though different, deformations of their centrosomal MTs. Those seen for the free-sliding model are associated with the bending of MTs that slide tangentially along the cortex (Fig. 4b), while those for the no-sliding model arise from buckling of MTs under compressive loads induced by polymerization (Fig. 5b).

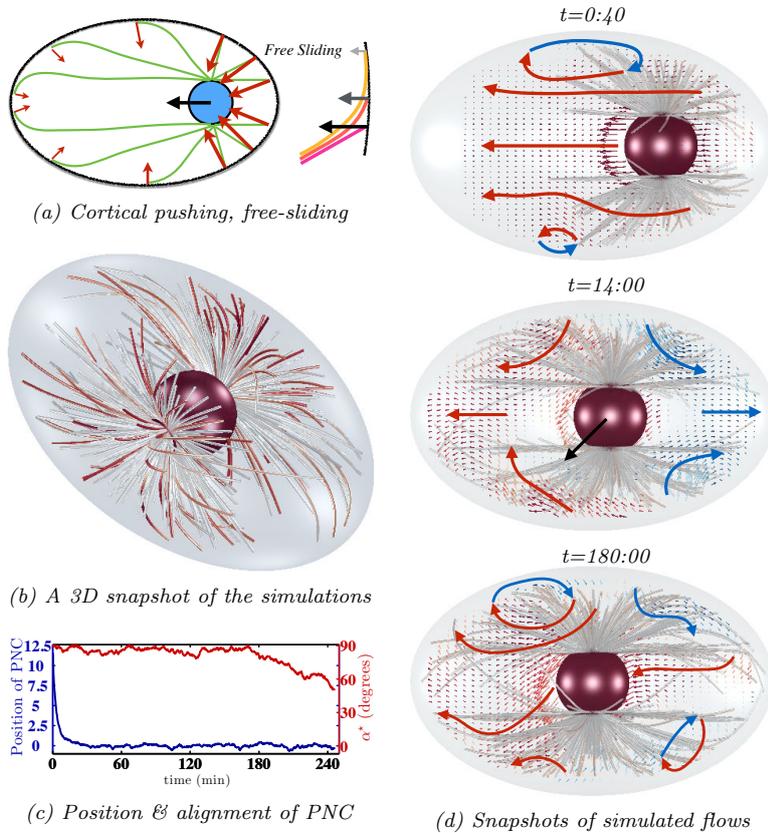

**Figure 4:** *(a) A schematic presentation of the cortical pushing free-sliding submodel. In this case the repulsive force is applied normal to the cortex and the magnitude of the force is reduces as MTs bend and become more aligned with the cortex; see §A. (b) A 3D snapshot of the simulation results. (c) The variations of PNC position and the angle between intercentrosomal axis and AP-axis with time. The PNC properly centers but fails to rotate in a physiologically reasonable time. (d) Snapshots of the cytoplasmic flows in the free-sliding submodel. These flows arise from the motion of the PNC and the astral MTs, the deformations of the MTs near the cortex. The cytoplasmic flow prior to centering is dominated by the motion of the PNC and the attached MTs.*

What underlies the differences, with respect to rotation, between these two models? In an idealized system where the PNC is pushed by an up-down symmetric set of astral MTs, the intercentrosomal axis would remain orthogonal to the AP-axis (i.e. $\alpha = \pi/2$) throughout the centering process (having started that way). However, a combination of dependency of polymerization forces on MT length[1] and the elongated shape of the cell periphery makes this orientation mechanically unstable and susceptible to the fluctuations, which MT dynamic instability can provide. Once seeded this "torque instability" produces a self-reinforcing, rotating torque upon the PNC moving it towards proper position, which is a mechanically stable equilibrium. This instability is discussed in detail in §A. The magnitude of this torque is determined by how much force is exerted on the MTs at the periphery and how quickly this force is transferred to the anchored minus-ends at the PNC, compared to the average time that MTs stay in contact with the cortex. In the no-sliding model, the polymerization forces scale with the buckling instability forces, $\pi^2 E/L^2$, and are applied largely in the direction of the fiber. Due to inextensibility of the MTs these forces are quickly (for straight MTs, immediately) transferred to the minus-end of the MTs. In the free-sliding model, the polymerization force pushing on MTs in contact with the cell periphery scales with the force required for bending, which is smaller than the buckling force. The time-scale for the bending forces to transfer this force is determined by the elastic relaxation time, $\tau^E = (8\pi\mu L^4)/(E\ln(\epsilon^{-2}e^{-1}))$; for a MT of length $L = 10\mu m$, $\tau \approx 40$ min. This time is much larger than the average time MTs interact with cortex ($1-10$ sec). The combination of weaker forces on the cortex and significantly slower force transfer result in a much smaller net "torque instability" and underlies the absence of PNC rotation in the free-sliding model. See §A for detailed analysis and direct quantifications of these two effects.

---

[1] The polymerization forces are balanced by the elastic deformation of the MTs resulting in $F^{\text{pushing}} \propto E/L^2$.

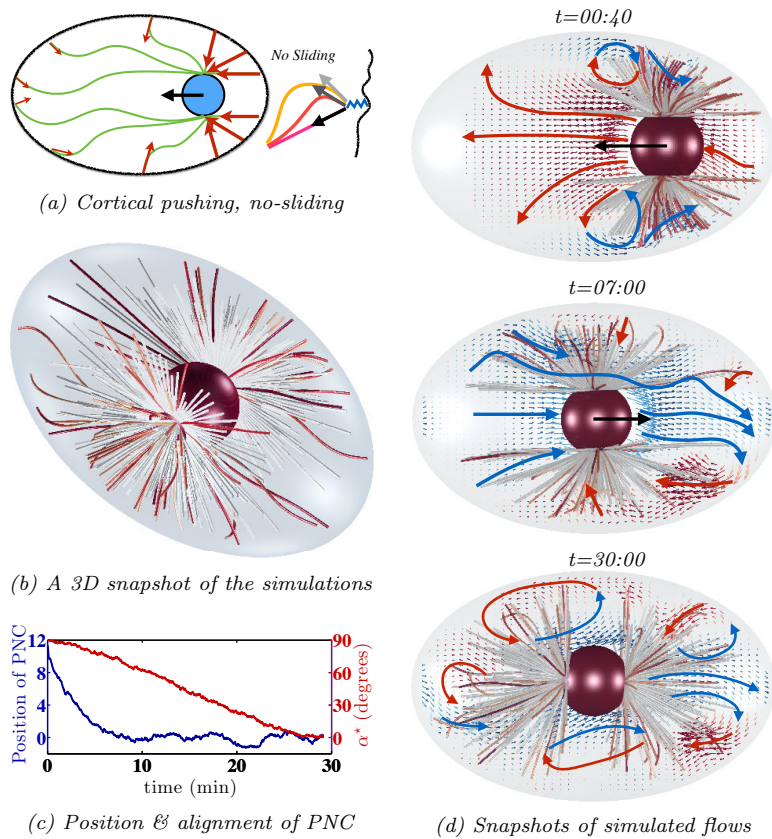

**Figure 5:** *(a) A schematic presentation of the cortical pushing no-sliding submodel. In this case, once the MTs reach closer than a minimum distance from the cortex, the position of their plus-end is fixed by attaching it to a stiff spring. The spring is released upon catastrophe; see §A. (b) A 3D snapshot of the simulation results. The MTs are buckled near the cortex. (c) The variations of PNC position and the angle between intercentrosomal axis and AP-axis with time. The PNC properly centers and aligns with the AP-axis within a physiologically reasonable time. (d) Snapshots of the cytoplasmic flows in the no-sliding submodel. The generated flows arise from the motion of the PNC and the astral MTs, and the active flows generated by buckling of the MTs near the cortex. These active flows are generally in the opposite direction of the polymerization of the MTs and remain as strong as the flow induced by the motion of the PNC throughout the entire migration process.*

We now discuss the generated flows of these two models, shown in Figs. 4d and 5d. Flows in both models are generated by the average motion of the PNC/MT-array complex (similar to the flow shown in Fig. 2b) and flows due to MT deformations near the periphery. The difference between the two cytoplasmic flows arises from the difference in the deformations of their MTs. In the free-sliding model, the end-forces acting on the MTs, as well as the strength of the flow generated by them, decrease as the MTs bend and align further with the cortex. Indeed, in the limit of MTs being aligned with the cortex the growth reaction does not generate any flow, and Fig. 4b shows that many MTs are aligned with the cortex in this model. As a result, unless the PNC is within 20% of the AP-axis length from the center, the flow induced by the average motion of the PNC/MT-array complex in the free-sliding model has a dominating effect and the flow remains qualitatively similar to Fig. 2b. However, upon approaching the center the PNC velocity vanishes and the weak flow induced by the deformations of the MTs near the periphery becomes comparable to the weakening flow induced by motion of the PNC/MT-arrays; a snapshot of the flow in this regime is shown in Fig. 4d (middle) where the arrows show the fluid flow streamlines which are in the direction of bending deformations and orthogonal to the cortex.

In the no-sliding model, the flow induced by MT motion near the periphery is comparable to that induced by the average PNC motion for most of the migration process. Unlike the free-sliding case, the flows near the periphery are now more in the direction of the MTs impinging upon it. Note that after the MTs reach the cortex, to continue the growth process they are pushed away from the boundaries with the same speed as they grow. Thus the generated flows near the cortex scale with the polymerization rate and are primarily in the direction opposite the polymerization direction (along the MT centerline).[2] Finally, the middle and bottom panels of Fig. 5d show that the same flow patterns are also observed near the periphery after completion of PNC centering and rotation.

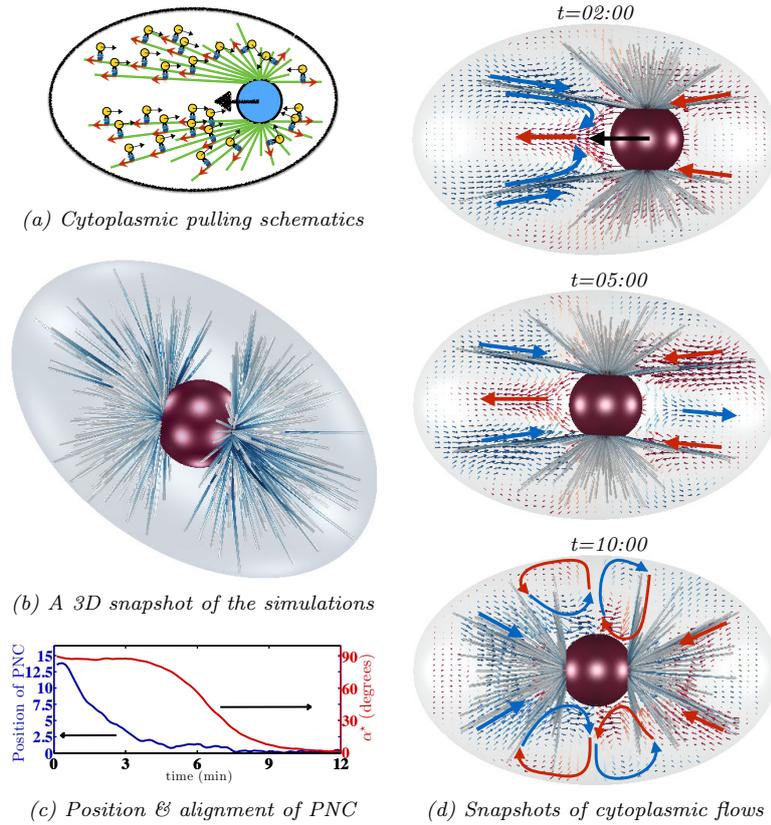

**Figure 6:** *(a) A schematic presentation of the cytoplasmic pulling model. In this model, cargo-carrying dyneins apply pulling forces on the MTs by walking and carrying cargos towards the minus-ends of MTs towards centrosomes. As a result of this translocation, the PNC migrates in the direction with the longest MTs; see §A. (b) A 3D snapshot of the simulation results. Since the MTs are under extensional load, they remain nearly straight. (c) The PNC position and the angle between intercentrosomal axis and AP-axis with time. The PNC properly centers and aligns with the AP-axis within a physiologically reasonable time. (d) Flow snapshots in cytoplasmic pulling model. These generated flows arise from the motion of the PNC and the astral MTs, and translocation of the cargo along the MTs. Unlike other mechanisms, (i) an active flow is generated in the opposite direction of migration on the anterior side of the cell, which is analogous to the flow of active puller swimmers [Saintillan and Shelley, 2013]; and (ii) the cytoplasmic flow is substantially stronger than the flow induced by motion of the PNC and astral MTs during and after completion of the migration process.*

Finally, a schematic of the cytoplasmic pulling model is shown in Fig. 6a. This model was initially proposed to explain observations in newly fertilized sand-dollar eggs and later investigated in an experimental study of [Kimura and Kimura, 2011] using *C. elegans*. This model was studied in our own earlier work on pronuclear migration [Shinar et al., 2011], where minus-end directed cargo-carrying dyneins walk on the MTs and so apply a pulling force on them towards their plus-end. Applying proper force balance [Shinar et al., 2011], the cargo exerts an equal and opposite force on the fluid. If dyneins are uniformly attached to the MTs, as is assumed here, the PNC complex moves in the direction of the longest MTs as these contribute the greatest pulling forces. See §A for details. As shown in Fig. 6c, this model can yield both centering and rotation on a time-scale that is comparable to experimental observations [Kimura and Onami, 2005].

---

[2]The generated flows near the cortex can be quite involved. The growth of the buckling amplitude, which is orthogonal to the direction of MTs, also contributes to the flow, as does the relaxation of the MTs switching to catastrophe.

In similarity to the cortical pulling model, the MTs are under extension and their deformations are small, which is evident in a 3D snapshot of the simulation illustrated in Fig. 6b. Unlike the cortical pushing and pulling models, which use the cell periphery as the mechanical substrate against which to exert forces on MTs, the force substrate is now the cytoplasm in which the cargos are immersed. Hence, the generated flow arises from two flows: (1) the flow induced by the average motion of the PNC/MT-array, and (2) the flow generated by the motion of cargos towards the minus-ends of MTs. Figure 6d clearly shows that the cytoplasmic flows thus produced are fundamentally different than those observed in the previous models. The key flow signature that is present in all the stages of migration is that, unlike the cortical pushing and pulling mechanisms, the cytoplasmic flow in the anterior is in the opposite direction of the motion of the PNC and is along the direction of cargo transport towards the centrosomes in the volume occupied by the astral MTs. These flows are illustrated by blue arrows in Fig. 6. The strength of the flow in the cytoplasmic pulling model is determined by the total active forces applied by the dynein motors, which in our model scales with the total length of the MTs. Consequently, the strength of flow throughout and after the migration process does not change significantly. On the other hand, the average velocity of the PNC monotonically decreases as the PNC approaches the center of the cell. Hence, as we see in Fig. 6d (middle and bottom panels), after centering the strength of cytoplasmic flows is much stronger than the flow near the PNC (which is determined by its small fluctuating velocity). This leads us to another key difference between the cytoplasmic pulling and the cortically-based models: in the cytoplasmic pulling model the ratio of the cytoplasmic velocity magnitude to that of velocity fluctuation of the PNC after centering is $\gg 1$ while for the other mechanisms this ratio is $\mathcal{O}(1)$.

There are general physical principles that underlie the gross cytoplasmic flows structures that we observe for the different models. First, the flows arising from cytoplasmic pulling resemble the average flows generated by "Puller" microswimmers [Saintillan and Shelley, 2013] such as Chlamydomonas reinhardtii [Drescher et al., 2010]. The reason lies in the ensemble behavior of the PNC/MT-array and the immersed dynein motors. As shown in Fig. 6a, for any force applied by the attached dynein motors on the MTs, there is an equal and opposite force is applied by the cargos to the cytoplasmic fluid. Thus, while the force on the PNC/MT-array is not zero, the net force on the PNC/MT-array and cytoplasmic fluid is identically zero. In particular, the motion of each cargo on MT-array generates a force-dipole, described by a tensor, whose symmetric part gives the net stress on the system, which unlike the net force is not zero. For straight MTs and a uniform distribution of the motors the net stress tensor induced by motor activity is

$$\sigma_{motor} = \frac{1}{2} \sum_{i=1}^{N} L_i \left( \mathbf{p}_i \mathbf{r}_i + \mathbf{r}_i \mathbf{p}_i \right) f_{motor},  \tag{1}$$

where $L_i$ and $\mathbf{p}_i$ are the length and unit tangent vector of the $i^{\text{th}}$ MT, respectively, and $\mathbf{r}_i$, taken as orthogonal to $\mathbf{p_i}$, gives the relative position of the cargo with respect to its attachment point on the $i^{th}$ MT. The antisymmetric part of the force dipole determines the net torque on the system which is also zero. Zero net force/torque and a finite *active* stress are the two hallmarks of active, self-propelled particles [Saintillan and Shelley, 2013]. As a result, the flow far from such particles is that generated by a *stresslet*. While in our system the size of the "active particle" — the PNC/MT-array — is similar to the scale of its confinement, the flow still closely resembles that generated by a Puller particle in an open flow (see Fig. 5 of [Saintillan and Shelley, 2013]).

The cortically-based models all generate grossly similar cytoplasmic flows (though different MT deformations) as they share the feature that the force is transduced to the PNC by MT pushing/pulling on/from an immobile cortex. In these conditions, the net force on the PNC/MT-arrays and the cytoplasm is non-zero. The cytoplasmic flows then resemble those generated by a force monopole, or a *Stokeslet*.

## 4 Discussion

We have presented a framework for simulating cellular assemblies that incorporates HIs amongst MTs and other bodies and surfaces while treating MTs as flexible fibers with appropriate dynamic instability kinetics.

This allows us to compute the cytoplasmic flows due to the motion of MTs and other immersed bodies within the cell. A more detailed discussion of the method is given in our concurrent work [Nazockdast et al., 2015]. While our approach is detailed, we have made several assumptions. While steric interactions of MTs with the cortex was modeled by a repulsive force, such interactions were not included between MTs; hence MTs can pass through one another. Fortunately this a very rare event in our simulations as the volume fraction of MTs is very low ($\phi \approx 10^{-5}$). We also assumed that MTs are mechanically *anchored*, say rather than hinged, to the PNC (the centrosome is not explicitly represented) at their minus-end. We know of no direct evidence to support, or confute, this assumption but the effect of other conditions can easily be explored in our framework. Experimental studies show that the PNC is very stiff [Dahl et al., 2004] and so the PNC is taken as a rigid sphere. Finally, we have used a continuum model for the force generated by cytoplasmic dyneins on the MTs, as in [Kimura and Kimura, 2011; Shinar et al., 2011]. We find that the number of dyneins per unit MT length giving reasonable migration times to be $n_{dyn} = 0.20(\mu m)^{-1}$, implying that many MTs will have no motor on them at any given time. Such discrete effects are not included in our continuum description. Generally, improving upon these assumptions will not change any of the conclusions drawn with regards to the effect of HIs and the generated flows in each model.

We showed also that previous approximations of the PNC drag coefficient, that either ignore HIs or include them partially, over-predict or under-predict active forces by an order of magnitude. For example, using such approximations [Reinsch and Gönczy, 1998] estimated that $\mathcal{O}(100)$ cortically-bound dyneins were sufficient to center the PNC/MT-array in *Xenopus* egg in physiologically reasonable times. However, our modeling shows that, for an analogous ratio of AP-axis length to average aster radius, confinement increases the drag by roughly 6-fold, and that $\mathcal{O}(600)$ are needed for centering in *Xenopus*. As another example, [Redemann et al., 2010] used the invaginations of softened cell membranes to estimate the number of cortical force generators at the end of anaphase. They find that 40 (80) or fewer MTs are being pulled towards the cortical anterior (posterior), with attachment times of approximately 25 seconds. This 1:2 ratio is consistent with estimates of [Grill et al., 2003]; and the modeling of [Kimura and Onami, 2007] assumes that this anterior contribution (unspecified in their text) remains fixed from migration to spindle elongation. In our simulations of the cortical pulling model, cortical attachment times are set to 20 seconds, and on average 32 MTs are simultaneously captured on each side after positioning (using $3pN$ stall force at each cortical capture site [Gönczy et al., 1999; Howard, 2001]).

We demonstrated the effect of confinement in reducing the PNC drag coefficients is significantly larger for translation than for rotation, giving that HIs cannot be simply described by an effective viscosity. Instead, the MT-arrays are more accurately described as porous layers around the PNC. In our computational model of the *C. elegans* embryo, the adimensional rotational and translational drag coefficients are close in magnitude and the PNC migrates twice its radius to the center. Thus the work required for rotation and translation of PNC are comparable. For larger cells, or for the smaller dimensions in consequent cell divisions, in *C-elegans* the translational drag coefficient might be substantially different from that the rotational. In a separate study, we numerically tested this by varying the length of the long axis from $60\mu m$ (60% of *C. elegans*) to $250\mu m$ while keeping the ratio of long to short axis fixed. Our results indicate that the translational drag initially increases with cell size of the cell, is maximized at $150\mu m$ and afterwards decreases a finite value. The rotational drag, however, increased super-linearly with the length of the AP-axis (to be published separately). This may have consequences on whether translation and rotation can be accomplished by a single force transduction mechanism on a single time-scale.

We studied the dynamics of PNC/MT-array positioning using simple instantiations of cortical pulling, cortical pushing, and cytoplasmic pulling mechanisms. These showed that cortical and cytoplasmic pulling can center and rotate the PNC within reasonable times. A cortical pushing mechanism can only properly position the PNC when MTs have *end-on* interactions with the cortex.

With regards to the cortical pushing mechanism, although *in vivo* observation in yeast and *in vitro* experiments on positioning of MT asters within microfabricated chambers show that polymerization forces can position the PNC or aster, respectively, there is little evidence that supports strong involvement of cortical pushing mechanisms in PNC migration of *C. elegans* embryo. In our simulations of a no-sliding pushing model that gave reasonable positioning time, the turnover time of MTs polymerizing against cortex was taken as $\tau = 5$ seconds, while *in vivo* observations suggest $\tau = 1-2$ seconds [Kozlowski et al., 2007]. Changing this parameter to $\tau = 3$ seconds increases the centering time to $\tau_c \approx 15$ minutes and rotation time $\tau_r \approx 45$ minutes (see §A) which is already much longer than experimental observations. This suggests that, at least in the first cell division, that cortical pushing is not the dominant mechanism for migration.

In our simulations of a cortical pulling model, based on that due to [Kimura and Onami, 2007], centering is only achieved in a very narrow range of parameters while rotation is robustly observed (see §A). A similar observation was reported by [Kimura and Onami, 2005] in their earlier modeling of pronuclear migration by cortical pulling forces. On the other hand, studies show that the PNC can rotate and properly align with the AP-axis in spherically shaped eggshells [Hyman and White, 1987; Tsou et al., 2002]. This cannot be achieved by cytoplasmic pulling, though can be achieved by cortical pulling where the requisite asymmetry is found in the distribution of cortical motors rather than in cell geometry.

The main experimental evidence in favor of the cytoplasmic pulling mechanism is given by [Kimura and Kimura, 2011], where they find a significant temporal correlation between the transport of cargo along MTs and PNC centration. Our simulation results (presented here and in [Nazockdast et al., 2015]) also suggest that between the three models of PNC migration, cytoplasmic pulling is the only model that robustly predicts centering and rotation of the PNC in physiologically reasonable times when parameters of the model are varied within a biologically feasible range. Other indirect evidence involves the observed streaming of vesicles and yolk granules along MTs towards the centrosomes during the migration process [Shinar et al., 2011]. This is in agreement with our simulated cytoplasmic flows. These observations suggest that either or both of the cytoplasmic and cortical pulling mechanisms could be involved in the pronuclear migration process, and bare observations of the dynamics of centering and rotation of PNC are insufficient to determine a specific mechanism.

As an alternative, we propose that the structure of cytoplasmic flows may select between the different possible active mechanisms involved in cellular processes such as pronuclear migration. We show, through simulation, that each of these force transduction mechanisms leaves its own specific fingerprint in the generated cytoplasmic flows; these features are directly related to the conformation of the MTs and how the force is transferred from the molecular motors and cell boundaries to the MTs. Since the flow signatures are generic features of each active mechanism, they can be used to study the possible contributions of these mechanisms in other stages of cell division. We are currently following up on these ideas by directly measuring the cytoplasmic flows by particle tracking methods.

We make a few final remarks on the importance of HIs in the dynamics of cellular structures. Although several studies discuss the importance of HIs in defining the dynamics of cilia and flagellar motion [Lauga and Powers, 2009], HIs between the fibers in cytoskeletal assemblies are almost entirely ignored. The assumption, here, is that the long-ranged HIs beyond nearest neighbor distances is screened and the dynamics of each individual fiber is dictated by its local interactions with the neighboring fibers. This is typically implemented through a modified drag on the fibers [Broedersz and MacKintosh, 2014]. Although this picture may give an acceptable description of the nearest neighbor mediated dynamics of individual fibers, the collective dynamics of cytoskeletal structures (PNC/MT-arrays here) could be incorrectly predicted. This is because the collective forces and stresses of the material can produce a large-scale background flow against which the microscopic elements evolve. In this instance, the inclusion of HIs leads to the interpretation of the MT-array as a porous shell, yielding a much more accurate representation of the dynamics of the entire PNC/MT-array structure. To summarize, HIs are an integral part of the mechanics of cellular assemblies and theoretical coarse-graining of their effect demands considerable care and attention to the relevant time and length scales. Dynamic simulation provides a powerful tool to investigate such relationships.

# Acknowledgment

We extend our thanks to Sebastian Füerthauer, Hassan Masoud, Tong Gao, Michael O'Neil, Jonathon Howard, and Carlos Garzon-Coral for helpful discussions. We acknowledge support from National Institutes of Health Grant 1R01GM104976-01.

# A  Supplementary Materials

## A.1  Numerical methods

In this section we give a brief description of the highly efficient computational method used to simulate the hydrodynamic interactions (through the cytoplasm) of MTs with each other, the pronuclear complex, and the cell cortex. This method also accounts stably for MT flexibility, their dynamic instability, and their interactions with molecular motors. Further details of the numerical method is given in [Nazockdast et al., 2015] (see also [Tornberg and Shelley, 2004] for earlier work). For cellular flows inertial effects can be safely ignored. While we assume the cytoplasm is assumed Newtonian [Daniels et al., 2006], the mechanical responses of the overall system of cytoplasm and cytoskeleton is in general non-Newtonian. The flow of a Newtonian cytoplasm is described by the incompressible Stokes equation:

$$\mu \Delta \mathbf{u} - \nabla p = \mathbf{0} \ \& \ \nabla \cdot \mathbf{u} = 0, \tag{2}$$

where $\mu$ is the bulk viscosity, $\mathbf{u}$ is the (cytoplasmic) fluid velocity, and $p$ is the pressure. We represent solutions to the Stokes equations using a boundary integral formulation [Pozrikidis, 1992] where the fluid velocity is represented as a distribution of fundamental solutions to the Stokes equations on all immersed and bounding surfaces. The densities of these distributions is determined by the application of boundary conditions, such as the no-slip condition (surface velocity is equal to fluid velocity). A boundary integral formulation reduces the computational problem from 3D (solving the Stokes equations in the fluid volume) to the 2D problem of solving coupled singular integral equations on all the immersed and bounding surfaces. In formulating the contributions from the bounding surface $S_E$ (the cortex) and any surfaces of internal bodies $S_I$ (here, only the PNC), we use a distribution of stresslets, a representation due to [Power and Miranda, 1987] which generates well-conditioned $2^{nd}$-kind Fredholm integral equations. The contributions due to the motion of MTs are be treated specially due to their slenderness, as their surface integrals can be reduced, through asymptotics, to integrals of Stokeslets along their center-lines.

To be specific, consider $N$ MTs attached to the pronuclear complex, with the entire structure contained within the cell cortex. The $n^{th}$ MT has center-line position $\mathbf{X}_n(s,t)$, where $s$ is arclength measured from the point of attachment and $0 \leq s \leq L_n$ with $L_n$ the MT's length. Then, the fluid velocity at a point $\mathbf{x}$ within the cytoplasm can be given as

$$\mathbf{u}(\mathbf{x}) = \mathbf{u}_{MT}(\mathbf{x}) + \mathbf{u}_E(\mathbf{x}) + \mathbf{u}_I(\mathbf{x}) \ \text{where} \tag{3a}$$

$$\mathbf{u}_{MT}(\mathbf{x}) = \sum_{m=1}^{N} \mathbf{u}_{MT,m}(\mathbf{x}) = \sum_{m=1}^{N} \int_0^{L_m} \mathbf{G}(\mathbf{x} - \mathbf{X}_m(s')) \cdot \mathbf{f}_m(s') ds' \tag{3b}$$

$$\mathbf{u}_E(\mathbf{x}) = \int_{S_E} \mathbf{q}_E(\mathbf{x}') \cdot \mathbf{T}(\mathbf{x} - \mathbf{x}') \cdot \mathbf{n}(\mathbf{x}') dS_{x'} \tag{3c}$$

$$\mathbf{u}_I(\mathbf{x}) = \int_{S_I} \mathbf{q}_I(\mathbf{x}') \cdot \mathbf{T}(\mathbf{x} - \mathbf{x}') \cdot \mathbf{n}(\mathbf{x}') dS_{x'} + \mathbf{G}(\mathbf{x} - \mathbf{x}_c) \cdot \mathbf{F}_I^{ext} + \mathbf{R}(\mathbf{x} - \mathbf{x}_c) \cdot \mathbf{L}_I^{ext} \tag{3d}$$

That is, the velocity is expressed as a sum of three sets of integrals, over MTs centerlines, the external boundary ($S_E$; the cortex), and internal boundaries ($S_I$; the pronuclear complex). Here $\mathbf{G}(\mathbf{r}) = (\mathbf{I} + \hat{\mathbf{r}}\hat{\mathbf{r}})/(8\pi\mu|\mathbf{r}|)$, with $\hat{\mathbf{r}} = \mathbf{r}/|\mathbf{r}|$, is the single-layer fundamental solution for the Stokes equations (the Stokeslet, a $2^{nd}$ rank tensor), $\mu$ is the fluid viscosity, and $\mathbf{f}_m$ is the force/length that the $m^{th}$ MT exerts upon the fluid; $\mathbf{T} = -3\hat{\mathbf{r}}\hat{\mathbf{r}}\hat{\mathbf{r}}/(4\pi\mu|\mathbf{r}|^2)$ is the double-layer fundamental solution (the stresslet, a $3^{rd}$ rank tensor), and $\mathbf{q}_{E,I}$ are vector densities to be determined [Power and Miranda, 1987]. Note that a distribution of stresslets on the surface of a body (immersed in a Stokesian fluid) produces identically zero net force and/or torque upon that body. Thus, in Eq. (3c), to account correctly for any applied forces and torques, explicit Stokeslet and Rotlet singularities are included inside the body. In particular, $\mathbf{F}^{ext}$ and $\mathbf{L}^{ext}$ are the external force and torque, respectively, exerted upon the PNC ($S_I$) where $\mathbf{R}(\mathbf{r}) \cdot \mathbf{L} = \mathbf{L} \times \hat{\mathbf{r}}/(8\pi\mu|\mathbf{r}|^2)$ is the *Rotlet* fundamental solution to the Stokes equations, and $\mathbf{x}_c$ is any point interior to $S_I$. Finally, the vector $\mathbf{n}$ is the unit outer normal to the $S_E$ and $S_I$ surfaces.

Taking the two limits $\mathbf{x} \to S_I$ or $S_E$ generates integral equations for the densities $\mathbf{q}_{E \text{or} I}$, respectively. Requiring that the outer boundary $S_E$ be stationary, or $\mathbf{u}(\mathbf{x}) = \mathbf{0}$ for $\mathbf{x} \in S_E$, generates the limiting integral equation

$$\mathbf{u}_{MT}(\mathbf{x}) + \mathbf{u}_I(\mathbf{x}) - 4\pi \mathbf{q}_E(\mathbf{x}) + \int_E \mathbf{q}_E(\mathbf{x}') \cdot \mathbf{T}(\mathbf{x}' - \mathbf{x}) \cdot \hat{\mathbf{n}}(\mathbf{x}') dS_{x'} = \mathbf{0}, \qquad (4)$$

Likewise, the limiting integral equation for $\mathbf{x} \in S_I$ is

$$\mathbf{U}_I + \mathbf{\Omega}_I \times (\mathbf{x} - \mathbf{x}_c) = \mathbf{u}_{MT}(\mathbf{x}) + \mathbf{u}_E(\mathbf{x}) + 4\pi \mathbf{q}_I(\mathbf{x}) + \int_I \mathbf{q}_I(\mathbf{x}) \cdot \mathbf{T}(\mathbf{x}' - \mathbf{x}) \cdot \hat{\mathbf{n}}(\mathbf{x}) dS_{x'} \qquad (5)$$
$$+ \mathbf{G}(\mathbf{x} - \mathbf{x}_c) \cdot \mathbf{F}_I^{ext} + \mathbf{R}(\mathbf{x} - \mathbf{x}_c) \cdot \mathbf{L}_I^{ext}.$$

In both Eqs. (4) & (5), the singular surface integrals are interpreted in the principal value sense.

One cannot take a limit of Eq. (3b) as $\mathbf{x}$ approaches a point on an MT. The resulting integral is undefined and the problem is instead treated through careful asymptotics [Götz, 2000; Johnson, 1980; Keller and Rubinow, 1976]. In this case it has been established that the velocity of the $n^{th}$ MT center-line, $\mathbf{U_n} = \partial \mathbf{X}_n / \partial t$, is given to leading order by

$$\mathbf{U}_n = \sum_{m=1}^{N} \mathbf{u}_{MT,m \neq n}(\mathbf{X}_n) + \mathbf{u}_E(\mathbf{X}_n) + \mathbf{u}_I(\mathbf{X}_n) + \left( \ln(\epsilon^{-2} e^{-1})/8\pi\mu \right) (\mathbf{I} + \mathbf{X}_{n,s} \mathbf{X}_{n,s}) \cdot \mathbf{f}_n(s), \qquad (6)$$

where we have assumed that arclength parameter is also a material parameter of the MT fiber which is a consequence of MT inextensibility. The subscript $s$ on $\mathbf{X}_{n,s}$ denotes a partial derivative with respect to $s$ (hence, $\mathbf{X}_{n,s}$ is the MT tangent vector), and $\epsilon = a/L_n$ is the MT aspect ratio. Here, we have neglected the velocity contribution from nonlocal self interactions of different segments of each individual fibers, which arise as a result of having curvature and non-uniform distribution of forces along the fiber. This interaction contributes to the velocity of the fiber to $\mathcal{O}(1)$ compared with the leading $\ln(\epsilon^{-2} e^{-1})$ term. Our numerical experiments [Nazockdast et al., 2015] show that including these nonlocal self interactions has negligible effect on the overall dynamics.

The force applied from an MT to the fluid, $\mathbf{f}$, is balanced with the hydrodynamic force from the fluid to the MT and is the sum of internal elastic forces and the forces applied by molecular motors or through interactions with the boundaries. External forces can be applied either at the MT ends (through boundary conditions), or along the length of the MT where $\mathbf{f} = \mathbf{f}^{elastic} + \mathbf{f}^{motor}$. The elastic forces are related to MT conformation through Euler-Bernouli beam theory by the constituitive relation $\mathbf{f}^{elastic} = -E\mathbf{X}_{ssss} + (T\mathbf{X}_s)_s$, with $E$ the MT flexural modulus, and $T$ is the MT's axial tension. The term $-E\mathbf{X}_{ssss}$ is the bending force per unit length and $(T\mathbf{x}_s)_s$ is tension force per unit length. The tension $T$ is determined by the condition of MT inextensibility [Tornberg and Shelley, 2004]. This constraint gives an auxiliary equation for tension by imposing $\mathbf{U}_s \cdot \mathbf{X}_s = 0$ which follows from differentiating the identity $\mathbf{X}_s \cdot \mathbf{X}_s = 1$ and interchanging $s$ and $t$ derivatives (that these derivatives can be interchanged follows from assuming $s$ gives a material parametrization of the MT position.

One method for modeling the (de)polymerization process is to discretely remove/add finite length segments from/to MTs in time. Our numerical experiments based on this approach shows that, to avoid numerical instabilities, extremely small time steps are needed. To overcome this limitation, we instead introduce a new parameterization variable, $\alpha$, of the MT centerline satisfying $\alpha = s/L(t)$ so that $0 \leq \alpha \leq 1$, and $L$ then appears explicitly in the reformulated equations. This removes the numerical instability and allows us to take much larger time-steps, now dictated by considerations of accuracy rather than numerical stability.

To fully determine the dynamics of the MTs and the PNC requires specifying $\mathbf{f}^{motor}$ which depends on the particular model of motor-protein type and activity. Three different models of PNC migration, resulting in different forms for $\mathbf{f}^{motor}$ are discussed in the next sections. Once $\mathbf{f}^{motor}$ is specified, Eqs. (5)-(6) are discretized using pseudo-spectral methods in space and an explicit/implicit backward time-stepping scheme. The latter treats the bending forces and tension implicitly (among other elements), which removes the high-order stability stiffness constraints from elasticity so that the time-step is chosen by the requirements of accuracy rather than numerical stability [Nazockdast et al., 2015]. This results in a linear system of equations to be solved at each time-step. For this we use GMRES [Saad and Schultz, 1986] with special purpose preconditioners [Nazockdast et al., 2015]. If each fiber is discretized by $M$ points, direct computation of the HIs between the $N$ MTs requires $(N \times M)^2$ operations which is very demanding for $\mathcal{O}(1000)$ or more MTs. We instead use a *Kernel-Independent* [Ying et al., 2004] implementation of the Fast Multipole Method (FMM) [Greengard and Rokhlin, 1987] to speed up the computation HIs, reducing the cost to $\mathcal{O}(N \times M)$. All the computations, including the FMM, are parallelized and scaleable.

## A.2 Biophysical models for positioning and their properties

### A.2.1 Cortical pushing

In the *cortical pushing* model, positioning is achieved by repulsive forces applied from the cortex when a growing MT reaches the cell boundary. The magnitude of the force is such that it stops the growth reaction process, or provides space for growth through moving the PNC away from the cortex or by deforming the growing fiber. *In vitro* measurements suggest that the stall force for growth process of the MTs is $F_P^S \approx 4.4$ pN [van Doorn et al., 2000]. If we take the average length of the astral MTs reaching the cortex throughout the centering process to be approximately $15\,\mu m$, the force threshold for a bucking instability is $F^B = \pi^2 E/L^2 \approx 0.4\,\text{pN} \ll 4.4\,\text{pN}$. Thus, in most interactions of MTs with cortex, MTs bend or buckle and continue the growth process rather than completely stalling it. Hence, larger forces are applied to the side of PNC with shorter MTs (larger buckling forces) resulting in the centering of the PNC.

We consider two variations of this model, each given schematically in Fig. 7. In the *free-sliding* model, we assume that the cortical repulsive forces are applied only along the normal direction to the surface i.e. the MTs cannot penetrate the boundary but they can grow or slide freely in the directions tangent to the boundary. We model this using a soft repulsive force, $\mathbf{F}^M = -F_P^S \exp(-\Delta r/\Delta r^\star)(\hat{\mathbf{n}}_{cor} \cdot \mathbf{x}_s)\hat{\mathbf{n}}_{cor}$ where $\Delta r$ is the minimum distance of the MT plus-end from the cortex. We set $\Delta r^\star = 0.02\mu m$ which allows us to simulate the dynamics using reasonable time-steps. When the MTs align tangentially to the boundary, $|\mathbf{x}_s \cdot \hat{\mathbf{n}}_{cor}| \ll 1$ resulting in vanishing end-point force on them.

In the *no-sliding* model, we constrain the sliding and growth along the boundaries and assume that the growing plus-ends of MTs remain fixed on the cortex as long as they are growing. We implement this constraint by a linear spring force at the attachment point: $\mathbf{F}^M = -K(\mathbf{X} - \mathbf{X}_{att})$, where $K$ is the spring constant (set to $100E/a_{pnc}^2$), and where $\mathbf{X}_{att}$ is the pinned plus-end position of the MT set by having reached a distance closer than $\Delta r^\star$ to the cortex. In both variations of the model, we use the empirical model given by [Dogterom and Yurke, 1997] based on their *in vitro* experimental studies, and relate the rate of growth to the applied end-force by $V_g = V_g^0 \exp\left(-(F^M \cdot \mathbf{x}_s)/F_P^S\right)$ where $V_g^0$ the growth rate under no compressive load. The *in vitro* measurements of [Janson et al., 2003] suggest that the turnover time of MTs in contact with cortex is proportional to their growth velocity. The measured rates of catastrophe in [Janson et al., 2003] are however generally smaller than $0.1\,\text{sec}^{-1}$ while *in-vivo* observations suggest $0.5\text{-}1\,\text{sec}^{-1}$ [McNally, 2013]. We use these observations and set the catastrophe rate to

$$f_{cat} = \max\left(f_{cat}^0 \frac{V_g^0}{V_g}, f_{min}\right) \tag{7}$$

where $f_{cat}^0$ is the rate of catastrophe under no compressive load. We include $f_{min}$ as the minimum allowed average rate of catastrophe in the model to incorporate the *in-vivo* observations. We changes this value from 0.1-0.33 sec$^{-1}$ to study its effect on the dynamics of migration, with the results presented in §A.5. Finally variations of $\Delta r^\star$ in the free-sliding model and spring stiffness, $K$, in the no-sliding model did not change the time-scales of PNC migration.

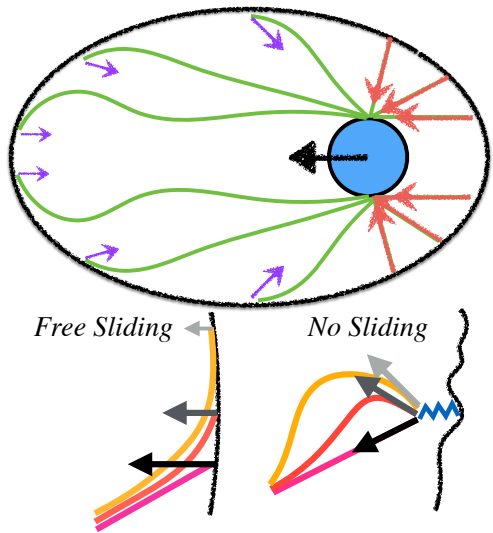

**Figure 7:** *A schematic of the cortical pushing model and its two variations: **free-sliding** where the MTs can grow and move tangential to the cortex. Polymerization forces are applied normal to the boundary; **no-sliding** where the plus end of the growing MTs remain attached to the cortex with a spring force.*

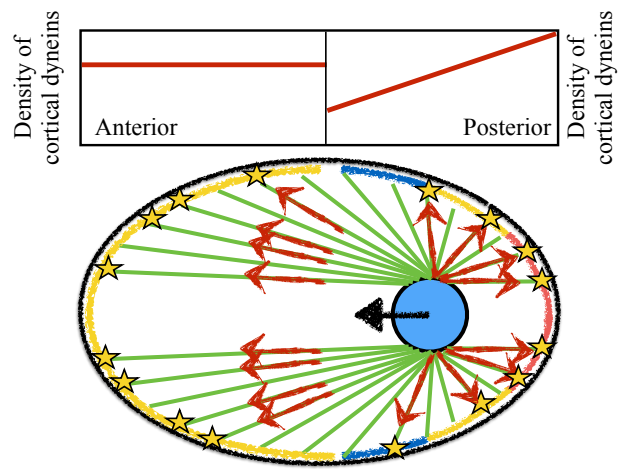

**Figure 8:** *A schematic of the cortical pulling model model; the stars represent the active cortical dyneins pulling on the MTs. The activated dyneins are uniformly distributed on the anterior side while on the posterior side their density decreases from the pole to the mid-plane. The blue, yellow and red colors on the boundaries denote, low, average and high number, respectively, of attached MTs.*

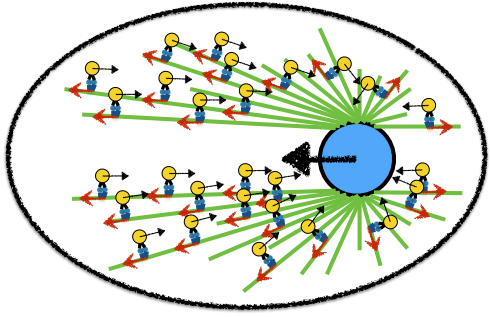

**Figure 9:** *A schematic of the cytoplasmic pulling model. More cytoplasmic dyneins are attached to the anterior MTs of longer lengths. As a result the PNC moves towards the anterior side until it centers. The cytoplasmic flows arise from the motion of the PNC and the astral MTs, and the motion of cargos towards the centrosomes.*

### A.2.2 Cortical Pulling

Our model for cortical pulling on the PNC is based on the asymmetric attachment of MTs to cortically bound dyneins on the anterior and posterior sides of the cell. It closely follows the model prosed by [Kimura and Onami, 2007], and is motivated by experiments of [Tsou et al., 2002] showing that LET-99 protein is highly enriched on the posterior side of the cortex in regions close to the mid-plane. LET-99 prevents MTs from attaching to cortically bound dyneins, and results in asymmetric pulling forces and a net force on the PNC towards the anterior. Following [Kimura and Onami, 2007], we assume that the rate of MT capture by cortically bound dyneins on the anterior cortex is constant, while the capture rate on the posterior side decreases linearly with distance along the $AP$ axis from the posterior pole to the PNC center; See Fig. sche-CorPull2. Also, similarly to [Kimura and Onami, 2007], we take the maximum rate of attachment on the posterior side to be $3/2$ times larger than on the anterior because the ratio of posterior to anterior pulling forces has been estimated to be $3/2$ by [Grill et al., 2003] during the spindle elongation process in anaphase. The slope of decrease in attachment probability is determined by imposing force balance at the PNC center.

### A.2.3 Cytoplasmic pulling

In the cytoplasmic pulling model, cargo-carrying dyneins walk towards the minus-ends of MTs and apply a pulling force on MTs that is equal and opposite to the hydrodynamic force needed to push the cargo through the cytoplasmic fluid. The basics are that a PNC starting migration on the posterior side of the cell has centrosomal MTs that can grow longer in the anterior direction than in the posterior. This leads to more attached cytoplasmic dyneins and payloads and hence a net force on the PNC towards the anterior [Kimura and Onami, 2005, 2007; Kimura and Kimura, 2011; Shinar et al., 2011]. The force that dyneins apply to MTs is related to their walking speed through a force-velocity relationship of a single motor as $F = F^{stall}\left(1 - \max(|V|, V_{max})/V_{max}\right)$ [Shinar et al., 2011] where $F^{stall}$, $V_{max}$ are the stall force and maximum walking speed of the motor. The force and velocity of the motor are also related by the drag coefficient of the cargo: $F = \gamma_{cargo}V$. Combining the two relationships gives $F^{motor} = F^{stall}V_{max}/\left(V_{max} + F^{stall}/\gamma_{cargo}\right)$. Here we have assumed that the PNC velocity is negligible in comparison to the walking speed of motor-proteins. If we take the cargo to be approximately a sphere with an average radius of $0.25\,\mu m$, we can compute the average force and velocity of the dynein motors that are $F^{motor} = 0.83F^{stall} \approx 0.91\text{pN}$ and $V = 0.2V_{max} = 0.4\mu\text{m s}^{-1}$. Assuming the migration process takes place in roughly 5min, the average velocity of PNC is $V_{PNC} \approx 0.065\mu\text{m s}^{-1}$, which is much smaller than the walking speed of the motors and does not play a determining role in the analysis given for force-velocity relationship of the motors. We assume a uniform probability of attachment of dynein motors on the MTs and model the force applied by the motors by a continuum model: $\mathbf{f}^{motor} = 0.83n_{dyn}F^{stall}\mathbf{X}_s$ where $n_{dyn}$ is the number of the attached dyneins per unit length and is the only fitting parameter in the simulations. The predicted centering and rotation times with the choice of $n_{dyn} = 0.1(\mu m)^{-1}$ (presented in this study) was found to be in the range of experimentally observed values that is approximately 5 min.

## A.3 A torque instability in cortical pushing models

We provide a simplified model to demonstrate the minimum physics required for observing the torque instability that arises in cortical pushing models and properly aligns the intercentrosomal axis with the AP-axis. A schematic of the model is given in Fig. 10a. Here, we assume the PNC is centered and that the intercentrosomal axis forms the angle $\alpha^\star$ with the AP-axis, and $N$ *straight* MTs are interacting with the cortex. We consider two models for the pushing force applied from the cortex :

$$\mathbf{F}^1(L) = -\lambda_1 E L^{-2} \hat{\mathbf{n}} \tag{8a}$$
$$\mathbf{F}^2(L) = -\lambda_2 E L^{-2} \hat{\mathbf{p}} \tag{8b}$$

where $L$ is the length of the MT, $\hat{\mathbf{n}}$ is the normal to the surface pointing outwards and $\hat{\mathbf{p}}$ is the unit alignment vector of the MT.

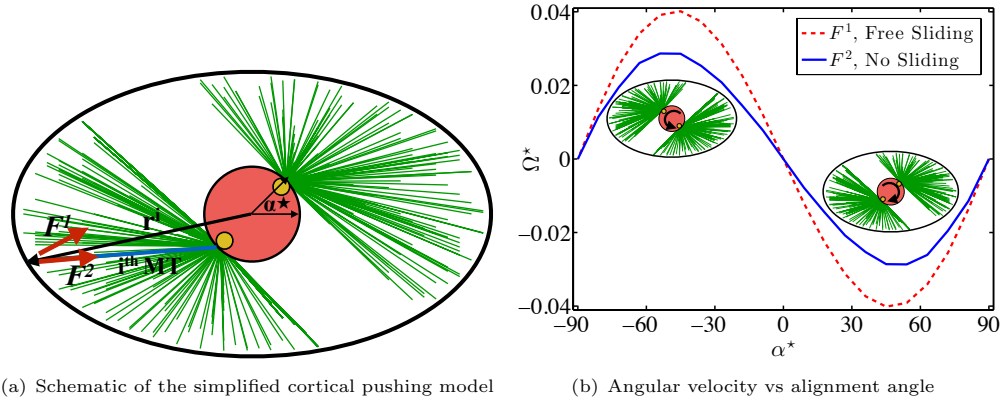

(a) Schematic of the simplified cortical pushing model    (b) Angular velocity vs alignment angle

**Figure 10:** *(a) A schematic presentation of the simple cortical pushing model used only to demonstrate the minimum physics required to achieve torque instability in cortical pushing mechanism. (b) The dimensionless angular velocity vs the angle of intercentrosomal axis with AP-axis ($\alpha^\star$).*

These two variations of force are simple representations of free-sliding (force normal to the cortex) and no-sliding (force in the opposite direction of the tangent of the MT). The net torque on the PNC is computed by summing the torque induced by the end-forces on the individual MTs:

$$\mathbf{T} = \sum_{i=1}^{N} \mathbf{r}_i \times \mathbf{F}_i(L_i)$$

where $\mathbf{r}_i$ is the vector connecting the center of the PNC to the plus-end of the $i^{th}$ MT. Note that this simple model does not contain the dynamic instability of MTs, any detailed hydrodynamic interactions (HIs), or MT flexibility. In both models we take the drag coefficient of the PNC and the attached MTs, $\gamma$, to remain constant with the change of $\alpha^\star$.

Figure 10b shows the dimensionless angular velocity, $\Omega^\star = \Omega/\Omega_0$, as a function of $\alpha^\star$ where $\Omega_0 = \frac{\lambda E N}{8\pi \eta a_{pnc}^4}$. The results clearly show that although the net torque is zero at $\alpha^\star = -90$ and $90$, infinitely small fluctuations induce a net torque that rotates the PNC towards the AP axis in a positive feed-back loop. On the contrary when the PNC is rotated away form the AP, a net torque is generated to realign the structure with the AP axis. Also the predictions from both variations of cortical forces are quantitatively close, which suggest that very slow rotation of the PNC in our detailed free-sliding model is not induced by the angle that the force is applied. Rather, as it is argued in the text, this contrast in behavior is induced by the smaller forces applied on the plus-ends as well as slower force transmission in the free-sliding model. These points are discussed in §A.4

## A.4 The PNC rotation dynamics in two cortical pushing sub-models

In the main text we discussed that the very slow rotation dynamics observed in the free-sliding sub-model of cortical pushing mechanism is a result of the combination of two effects: (1) In the free-sliding sub-model the MTs are bent, while in the no-sliding case MTs buckle, which requires much larger compressive forces than for bending deformations; and (2), the transfer of force from the plus-ends to the MT anchoring sites on the PNC (at minus-ends) are slower when the MTs undergo bending deformations.

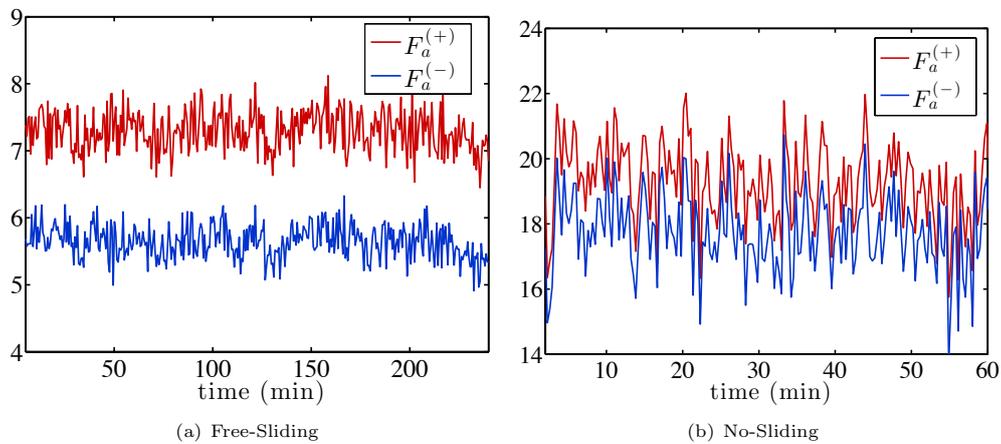

**Figure 11:** *The time-variations of the total magnitude of forces acting on the plus-ends, $F_a^{(+)}$, and the minus-ends, $F_a^{(-)}$, of the MTs for the (a) free-sliding and (b) no-sliding sub-models of cortical pushing mechanism. The red color shows the plus-end polymerization forces arising from interaction with the cortex and the blue color shows the forces on the minus-ends (anchoring sites).*

To test our hypothesis, we compute the total magnitude of forces on MT plus-ends that are interacting with the cortex, as well as the total force at the anchoring points of those MTs (minus-ends), as a function of time and for both free-sliding and no-sliding conditions. The results are presented in Fig. 11. The time-averaged total force on the plus ends, $F_a^{(+)}$, is 18 pN for the free-sliding case and 48 pN for no-sliding. This difference is more pronounced for the total force magnitude on their minus-ends which is 14 pN and 44 pN (more than 3 times larger) in the free- and no-sliding sub-models, respectively. A direct measure of how efficiently (fast) the force is transmitted from the plus- to minus-ends can be obtained by taking the ratio of the total force on the plus- and minus-ends. This ratio is 0.92 for the no-sliding sub-model while it is 0.75 for the free-sliding case. These observations confirm our hypothesis that the force applied on MT plus-end is substantially larger in the no-sliding model compared with free-sliding, and that the force transmission is less efficient in the free-sliding model compared with the no-sliding model.

## A.5 The effect of the rate of catastrophe of MTs interacting with the cell cortex on the time-scale of migration in the no-sliding cortical pushing model

We only consider the no-sliding model, since it is the only model (between the two) that correctly aligns the intercentrosomal axis with the AP-axis within a physiologically reasonable time. Figure 12 shows snapshots of simulations of the no-sliding cortical pushing model, using three values of minimum allowed catastrophe rates in Eq. (7), $f_{min} = 0.33, 0.2$ and $0.10$ (sec)$^{-1}$ respectively. For convenience we define the maximum turnover time of the MTs on the boundary as the inverse of the minimum rate of catastrophe, $\tau_c = 1/f_{min}$, and discuss the results in terms of this quantity. Thus the figures respectively correspond to $\tau_c = 3, 5$ and $10$ seconds. As expected, the MTs' deformation is increased with increasing the turnover time. The simplest interpretation of the dynamics would suggest that the total polymerization force applied from the cortex to the plus-end of MTs would be proportional to the turnover time. To examine this hypothesis we compute the of position and the angle of the centrosome poles with respect to the AP axis ($\alpha^\star$) as a function of time for three values of $\tau_c$ in Fig 13.

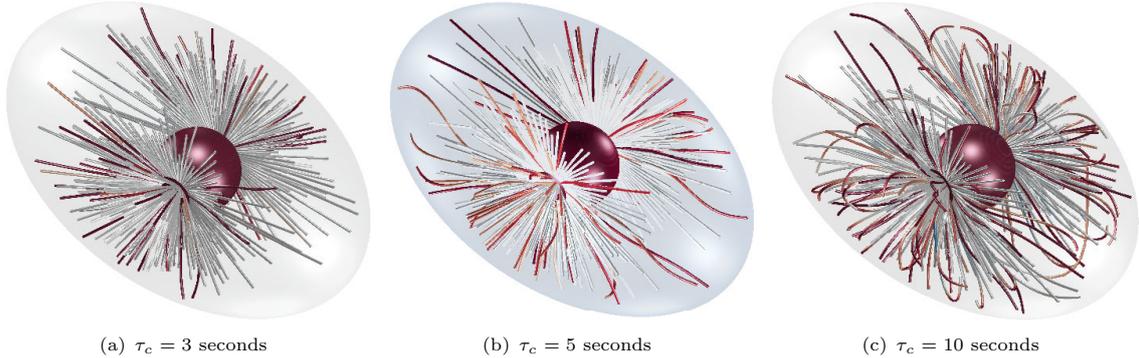

(a) $\tau_c = 3$ seconds      (b) $\tau_c = 5$ seconds      (c) $\tau_c = 10$ seconds

**Figure 12:** *Snapshots of the simulations of the no-sliding cortical pushing model at turnover times, $\tau_c = 3, 5$ and $10$ seconds, respectively. The colors denote the local tension of the MTs and they change from blue, denoting the MTs being under extension, to white, denoting no tension, and red, denoting the MTs being under compression.*

Comparing the time-scales of centering for $\tau_c = 5$ and $\tau_c = 10$ seconds shows that, although the initial migration speed is larger in $\tau_c = 10$ seconds, the centering time (taken here as the first time the PNC passes the center) is quantitatively similar (roughly 8 minutes) for both turnover times. On the other hand the centering time for $\tau_c = 3$ seconds is roughly doubled to 15 minutes. We relate this trend to the extent of the deformation of MTs as follows. Once the MTs are buckled (or generally deformed) beyond a threshold, the force it takes to continue the buckling process is substantially reduced. This is due to the fact that the deformation of a buckled MTs is a combination of bending and buckling modes of deformation. Since the bending is substantially less than the force producing a buckling instability, the continuation of the bucking would not result in a substantial increase in the time-average polymerization forces that are applied to the plus-end of MTs.

Another change in the dynamics due to the extent of deformation is in the time it takes the force to be transmitted from the plus-end of MTs to the anchoring sites. When the MTs is only mildly buckled, the force is transferred almost instantaneously from the plus to minus-ends (due to the extensibility of the MTs). However when those MTs are substantially buckled, at least the force due to the bending mode of deformation of the end-force is transmitted in the time-scale of the elastic relaxation time of the fiber which for MTs in *C elegans* is in the range of 40 minutes. A combination of these two effects result in the saturation of the speed of centering with the turnover time at $\tau_c = 10$ seconds.

The rotation dynamics of the PNC, however, follows a different trend. We see that the rate of rotation at $\tau_c = 10$ seconds is about 2 and 3 times faster than $\tau_c = 5$ and 3 seconds, respectively. The ratio of rotation to centering times are roughly 3, 3.5 and 1.6 in $\tau_c = 3, 5$ and $10$ seconds.

To test our hypothesis on the effect of deformation of polymerization forces, we compute the net force on the plus-ends interacting with the cortex as well as the net force on MTs' minus-ends anchored to the PNC for these three turnover times. Fig. 14 shows the total magnitude of the plus- and minus-end forces ($N_{MT} = 600$) as a function of time. As it can be seen, the total force acting upon the plus-ends (averaged over time) is increased from 14.8 to 19.3 and 23.3 pN for $\tau_c = 3, 5$ and 10 seconds. This trend shows that the magnitude of the end forces are only weakly proportional (sub-linearly) to the turnover time of MTs interacting the cell boundaries, and seem to be saturating at the largest $\tau_c$. This shows that once the MTs are buckled, the force required to continue the deformation is substantially reduced.

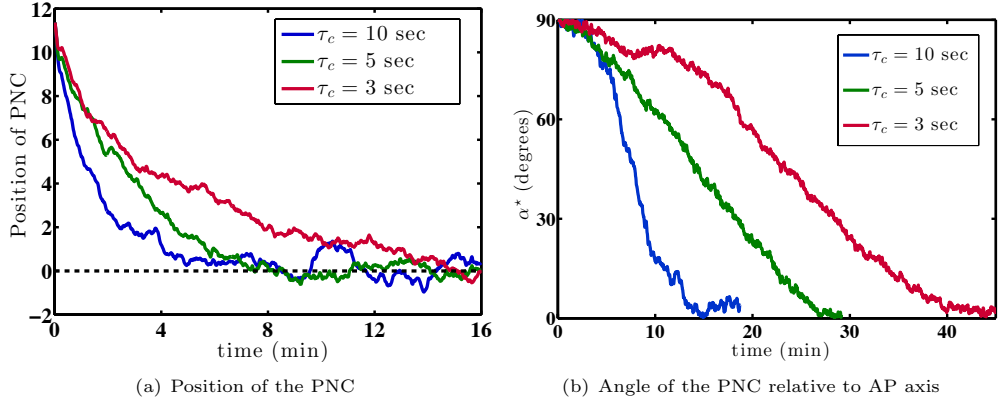

(a) Position of the PNC

(b) Angle of the PNC relative to AP axis

**Figure 13:** *(a) The position of PNC as a function of time for turnover times $\tau_c = 3, 5$ and $10$ seconds. (b) The relative angle of the intercentrosomal axis with AP-axis vs time for $\tau_c = 3, 5$ and $10$ seconds.*

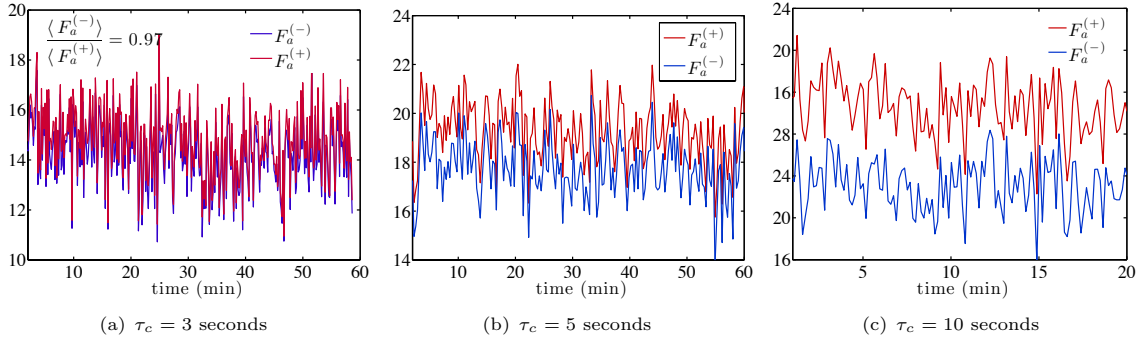

(a) $\tau_c = 3$ seconds

(b) $\tau_c = 5$ seconds

(c) $\tau_c = 10$ seconds

**Figure 14:** *The time-variations of the total magnitude of forces acting on the plus-ends, $F_a^{(+)}$, and the minus-ends, $F_a^{(-)}$ of the MTs at turnover times $\tau_c = 3, 5$ and $10$ seconds. The red color corresponds to the plus-end polymerization forces arising from interaction with the cortex and the blue color shows the forces on the minus-ends (anchoring sites).*

The total forces on the anchoring sites of the PNC also increase from 14.3 to 17.7 and 19.3 pN. Thus the net forces on the minus-ends for $\tau_c = 5$ and $\tau_c = 10$ seconds are quantitatively similar, resulting in similar time-scale for centering. The *effectiveness* of force transfer from the plus- to minus-ends of MTs can be evaluated by comparing the ratios of the net plus- to minus-end forces at the given turnover times. These ratios are 0.97 to 0.92 and 0.82 for $\tau_c = 3, 5$ and 10 seconds. We see that the ratio is very close to 1 for $\tau_c = 3$ and 5 seconds, demonstrating that the entire force is almost immediately transferred from plus- to minus-ends. This ratio is reduced for $\tau_c = 10$ seconds indicating a slower rate of force transfer.

## A.6  Cytoplasmic flows are generic features of active mechanisms

To demonstrate that the observed flow signatures are generic to each active mechanism, we present the results of our simulations for different values of of biophysical parameters that the ones presented in main text. We start by varying the rate of catastrophe in cortical pushing model. In Fig 15 we present the snapshots of the cytoplasmic flows in the no-sliding sub-model in the early stages of migration (top), after centering (middle) and after rotation (bottom) for $\tau_c = 10$ seconds (twice as much as the value used in the main text). To enable direct comparison we re-represent the results of the main text at $\tau_c = 5$ seconds.

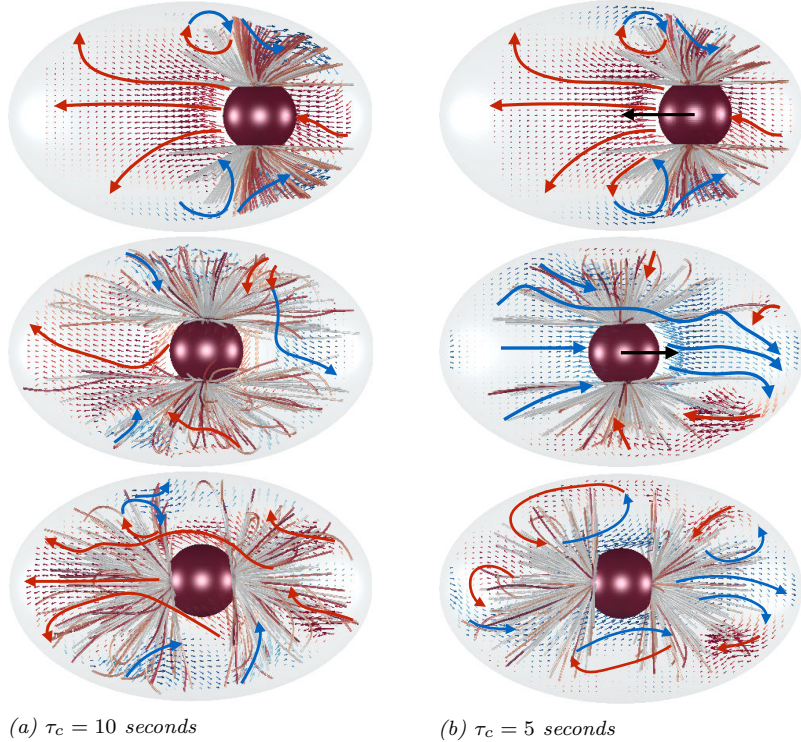

(a) $\tau_c = 10$ seconds    (b) $\tau_c = 5$ seconds

**Figure 15:** *Snapshots of the simulated cytoplasmic flows in the initial stages of migration, after centering, and rotation of PNC in cortical pushing with the no-sliding motion sub-model for two different turnover time of the MTs growing against the cortex : (a) $\tau_c = 10$ seconds, and (b) $\tau_c = 5$ seconds.*

The snapshots clearly show that as a consequence of increased turnover time the MTs are substantially more deformed at $\tau_c = 10$ seconds compared with $\tau_c = 5$ seconds. This, as we have showed in §A.5, also results in faster rotation of the PNC by at least a factor of 2. Despite these evident changes in the dynamics of the PNC migration and the shape of the MTs, the generated flow signatures remain unchanged with increase of the turnover time. Recall that these features for no sliding sub-model are strong flows near the cortex in the direction of the MTs and in the opposite direction of polymerization.

# B  Centering through the cortical pulling mechanism depends on the choice of model parameters

As mentioned in the main text, unlike the other two mechanisms, in the cortical pulling model the pronuclear centering is only achieved in a narrow range of parameters of the model used in this study. These parameters include the total number of cortically bound dyneins, the rate of detachment of captured MTs from cortically bound dyneins, and the slope by which the density of the cortically bound dyneins on the posterior side decrease from the pole to the dividing plane of the cell. Here we give few examples of failed centration of the PNC.

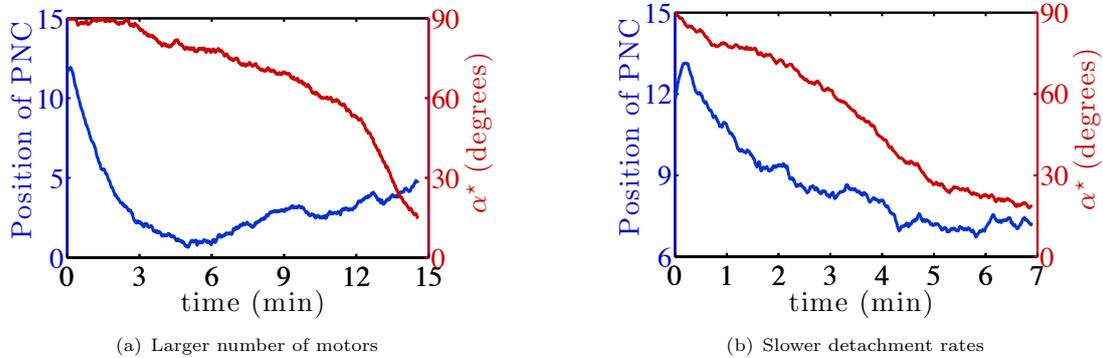

(a) Larger number of motors

(b) Slower detachment rates

**Figure 16:** *Two cases where the cortical pulling model does not center the PNC: (a) the number of cortical dynein motors are doubled with respect to the case of properly centered PNC in the main text; and (b) the rate of detachment is increased a factor of 2.*

For example, when the number of the cortically bound dynein motors is increased from 100 to 200, the PNC initially migrates towards the center. However, at later times the PNC moves back to the posterior side of the cell. Despite failure of centering, the intercentrosomal axis of PNC rotates and properly aligns with the AP-axis. These results are presented in Fig 16a.

In another example shown in Fig 16b, we increase the rate of detachment from 0.75 (1/min) to 1.5 (1/min). While the intercentrosomal axis successfully aligned with the AP axis, the PNC remained in the posterior side of the cell.

Changing the spatial variation of the cortical dyneins also resulted in failed centering, and successful rotation which are not shown here. It seems that this particular choice of model for cortical pulling mechanism allows multiple statistically stable solutions subjected to zero net force on the PNC, while other active mechanisms are not present. Our results suggest that the final position of the PNC depends critically on the competition between the different time-scales in this problem, which include the average time it takes MTs to grow to the cortex, the time the MTs remain attached to the cortex, and the time scale of migration.